\documentclass[12pt]{article}
\usepackage{graphicx, amsmath,amsfonts,amsthm, euscript,amscd}
\usepackage{braket}
\usepackage{tikz}
\usetikzlibrary{decorations.pathreplacing}
\usepackage{dsfont}
\usepackage{subfigure}
\usepackage[shortlabels]{enumitem}

\graphicspath{ {./images/} }
\usepackage{verbatim}
\usepackage{color}
\usepackage{ulem}
\newtheorem{theorem}{Theorem}[section]

\newtheorem{definition}{Definition}[section]

\newtheorem{remark}{Remark}[section]
\newtheorem{example}{Example}[section]

\usepackage{graphicx, amsmath}
\usepackage{color}
\definecolor{Blue}{rgb}{0.3,0.3,0.9}

\begin{document}

\normalsize

\title{Time-inhomogeneous Quantum Walks with Decoherence on Discrete Infinite Spaces}

\maketitle

\vskip 6pt
\author{ Chia-Han Chou and  Wei-Shih Yang}
\vskip 3pt
\indent Department of Mathematics, Temple University,\\
   \indent     Philadelphia, PA 19122

\vskip 6pt
Email: chia-han.chou@temple.edu, yang@temple.edu

\vskip 12pt


\vskip12pt
AMS classification Primary: 82B10


\begin{abstract}
In quantum computation theory, quantum random walks have been utilized by many quantum search algorithms which provide improved performance over their classical counterparts. However, due to the importance of the quantum decoherence phenomenon, decoherent quantum walks and their applications have been studied on a wide variety of structures. Recently, a unified time-inhomogeneous coin-turning random walk with rescaled limiting distributions, Bernoulli, uniform, arcsine and semi-circle laws as parameter varies have been obtained. In this paper we study the quantum analogue of these models. We obtained a representation theorem for time-inhomogeneous quantum walk on discrete infinite state space. Additionally, the convergence of the distributions of the decoherent quantum walks are numerically estimated as an application of the representation theorem, and the convergence in distribution of the quantum analogues of Bernoulli, uniform, arcsine and semicircle laws are statistically analyzed.

\end{abstract}

\section{Introduction}
\setcounter{equation}{0}

In classical computer science, researchers started utilizing randomness techniques such as Ulam and von Nuemann’s Markov Chain Monte Carlo (MCMC) method \cite{MetroUlam} to develop more efficient algorithms for tackling a wide variety of problems in 1940s. This method was later refined as the well known Metropolis-Hastings algorithm \cite{GantertAnnealing}, \cite{Hastings} with applications in different disciplines. The key idea behind these methods was that the true solution can be approximated with high probability by repeating Monte Carlo simulations.

In quantum computation, "qubit" takes a complex unit instead of "bit". In this setting, to preserve a cohesive quantum system, the family of qubits goes through unitary evolution instead of traditional gates in classical computation. After the completion of each algorithm, the state of the quantum system can be measured and observed, and the system collapses to one unique state from a superposition of various states. In this case, the probability of observing any given state after observation is proportional to the absolute value squared of the amplitude of the system at that state. Therefore, a false solution may be observed which is similar to Monte-Carlo methods. 
%However, if the algorithm is cleverly constructed, the correct solution is observed with significant likelihood.

New types of quantum  algorithms have been discovered since 1990s, and quantum mechanical properties give them better efficiency compared to existing classical algorithms. For example, both integer factorization and discrete logarithms
undergo an exponential speedup using Shor’s algorithm \cite{Shor}, and Grover’s search algorithm provides a quadratic speedup over any known classical search algorithm \cite{Grover}. Particularly on a discrete space, Grover’s algorithm is defined by discrete-time quantum walk, which is the natural extension of the quantum analogue of classical random walk on discrete spaces.

On the other hand, to manipulate or investigate a quantum system, it is not possible to keep a quantum system indefinitely isolated, coherent quantum system. However, if it is not perfectly isolated, for instance during a measurement, coherence is shared with the environment and appears to be lost with time and we call this phenomenon quantum decoherence. This concept was first introduced by H. Zeh \cite{ZehDecoh} in 1970, and then formulated mathematically for quantum walks by T. Brun \cite{BrunDecoh}. For both coin and position space partially decoherent Hadamard walk with strength $0<p\leq 1$, K. Zhang proved in \cite{ZhangQuantumLimit} that with symmetric initial conditions, it has Gaussian limiting distribution. More recently, the fact that the limiting distribution of the rescaled position discrete-time quantum random walks with general unitary operators subject to only coin space partially decoherence with strength $0<p< 1$ is a convex combination of normal distributions under certain conditions is proved by S. Fan, Z. Feng, S. Xiong and W. Yang \cite{YangConvexNormal}. Moreover, the time-inhomogeneous decoherent quantum analogues of Markov chains on finite state spaces were also studied and proved their equilibrium properties by C. Chou and W. Yang \cite{ChouQMarkov}. The decoherent quantum analogues random walks on discrete infinite space will be defined and elaborated in this paper.

In optimization problems, simulated annealing is not only a probabilistic method of approximating the global maximum or minimum of a given function but also one of the applications of time-inhomogeneous Markov chains (details discussed in section 3 of \cite{GantertAnnealing}). Indeed, this method is inspired by the annealing procedure of the metal working introduced by Kirkpatrick et al. \cite{KirkpatrickAnnealing} in 1980s.

Despite classical homogeneous Markov chain limit theorems for the discrete time walks are well known and have important applications in related areas \cite{Rurrett} and \cite{Mixing}, J. Englander and S. Volkov recently considered "Turning a Coin over Instead of Tossing It" \cite{EnglanderClassic} which is a time-inhomogeneous model defined as follows. Let $0<p_n<1$ be a sequence of deterministic numbers. Let $Y_n= -Y_{n-1}$ with probability $p_n$ and $Y_n= Y_{n-1}$ with probability $1-p_n$. Unlike i.i.d. increment simple random walk, this model has a very rich structure, for $p_n=\lambda n^{-\zeta}$, they obtained limiting distribution of $(Y_1+...+Y_n )/ N^{(1+\zeta)/2}$, and obtained Gaussian when $\zeta<1$, and uniform, semicircle and arcsine laws when $\zeta=1$ and Bernoulli when $\zeta>1$ among the scaling limits. 

Previously in 1950s, R. Dobrushin \cite{DobrushinInhomoMarkov} studied time-inhomogeneility on classical Markov chains and proved a definite central limit theorem in his thesis which provides the statement that, after centering and normalizing
with the standard deviation, the limit is standard normal. However, Dobrushin's theorem only applies to the case
when $0 < \zeta < 1/3$. Later in 2000s, Z. Dietz \cite{DietzInhomoMarkov} and S. Sethuraman \cite{SethuramanInhomoMarkov} obtained their scaling limits as symmetric beta distribution when $\zeta=1$ and the weak law of large numbers when $0<\zeta<1$ (strong law only when $0<\zeta<1/2$).

These classical results are naturally related to recently studied quantum random walk under decoherence. For Hadamard walk, when decoherence strength equal to 1, it is classical simple random walk. However, for an asymmetric coin operator when decoherence strength equal to 1, the resulting classical walk is not a sum of i.i.d. random variables; it is exactly the classical coin turning process with $\zeta=0$ in  \cite{EnglanderClassic}. The primary goal of this body of work is to consider the quantized version of the coin turning model with general parameters $\lambda, \zeta$ and the decoherence strength $0<p<1$.

We study a new model time-inhomogeneous quantum walk with decoherence on discrete infinite spaces, and obtain a representation theorem for time-inhomogeneous quantum walk through path integral expressions.  As the applications of the theorem, we introduce a new quantum algorithm with Monte Carlo technique to numerically approximate not only the classical symmetric Beta distributions and Bernoulli distributions, but also the limiting distributions of the decoherent time-inhomogeneous quantum walks.

Lastly, motivated by N. Konno with his obtained scaling limit of pure quantum Hadamard walk as the quantized normal distribution in \cite{KonnoHadamardLimit}, we introduce the quantum analogues of the classical distributions: arcsine, uniform, semicircle, and Bernoulli laws by considering the coherent inhomogeneous quantum walk in infinite discrete spaces. Even though it is extremely hard to study analytically the quantized classical distributions, we study their scaling limits and convergent rates, compare with their classical analogues, and statistically conclude that the quantum analogues of Bernoulli and Beta distributions under appropriate scaling exponents in this case are Bernoulli Laws.  

This paper is organized as follows. In Section \ref{sec:notation}, we set the notations, definitions and introduce the model. In Section \ref{QRW}, we develop a path integral formula for time-inhomogeneous decoherent quantum random walks and obtain a representation theorem for decoherent time-inhomogeneous random walks, Theorem \ref{RepreQuant}. In Section \ref{ApplicationRW}, we present the applications of the presentation theorem. In Sections \ref{QCD} and \ref{sec:quantumBernoulli}, we present the quantized classical distributions and numerically analyze their convergence and limiting distributions. In Section \ref{section:conclusion}, we make our conclusions and discuss some problems for further study.

\section{Quantum walks on $\mathbb{Z}$}\label{sec:notation}
In classical probability, a random walk on $\mathbb{Z}$ is a Markov process described by a
stochastic transition matrix T. On the other hand, for a quantum walk, instead of the transition matrix, the evolution of the system is described by a unitary operator U acting on a Hilbert space H. Several different models for quantum walks have been popularized.
The two most commonly used are the coined walk of Aharonov et al \cite{AharonovQuanGraph} and the quantum markov chain of Szegedy \cite{SzegedyMarkovC}. Recently, S. Fan, Z. Feng, S. Xiong and W. Yang et al. \cite{YangConvexNormal} demonstrated that under certain conditions, the limiting distribution of the rescaled discrete-time coin-space decoherent quantum walks is a convex combination of normal distributions. 

First, we consider a pure quantum random walk on the 1-dimensional integer lattices $\mathbb{Z}$. The state space is a Hilbert space $H=H_{p}\otimes H_c$, where  $H_{p}$ denotes the position space, and  $H_{c}$ denotes the coin space, and $\otimes$ is the tensor product. The orthonormal basis of the position space are $\{ |x\rangle, x\in Z\}$ and,  the orthonormal basis of the coin space are $|1\rangle$ and $|2\rangle$.

\begin{definition}\label{standOp}
	The standard  shift operator  $S: H \to H$ is a linear operator  defined as follows
	
	$$S(|x\rangle \otimes |1\rangle) =|x+1\rangle \otimes |1\rangle$$
	$$S(|x\rangle \otimes |2\rangle) =|x-1\rangle \otimes |2\rangle$$
	
\end{definition}

%\begin{definition}\label{flipOp}
%	The flip-flop shift operator  $S_f: H \to H$ is a linear operator  defined as follows
	
%	$$S_f(|x\rangle \otimes |1\rangle) =|x+1\rangle \otimes |2\rangle$$
%	$$S_f(|x\rangle \otimes |2\rangle) =|x-1\rangle \otimes |1\rangle$$
	
%\end{definition}

Note that $S$ can also be decomposed as following 

$$S=S^{+} \otimes |1\rangle\langle 1|+S^{-} \otimes |2\rangle\langle 2|$$
%$$S_f=S^{+} \otimes |2\rangle\langle 1|+S^{-} \otimes |1\rangle\langle 2|$$

where $S^{+}, S^{-}: H_p \to H_p$ are linear operators defined by 
\begin{eqnarray}
S^{+}(|x\rangle) :=|x+1\rangle \\
S^{-}(|x\rangle ) :=|x-1\rangle .
\end{eqnarray}

Let $F: H \to H$  be a unitary transformation on $H$ defined by 
\begin{eqnarray}
F=\sum_{x \in Z} |x\rangle \langle x|\otimes C,\label{eq:coin def}
\end{eqnarray}
where $C:H_c \to H_c$ is a unitary operator. We can observe here that $|x\rangle \langle x|$ is the projection operator over the position space $H_p$.

\begin{definition}\label{homoQRW}
	The evolution operator of standard quantum random walk is given by
	\begin{eqnarray}
	U=SF,
	\end{eqnarray}
	where $S$ is the standard shift operator. 
\end{definition}

Let $|\psi _0\rangle  \in H$. Then $|\psi _n\rangle =U^n| \psi _0\rangle = U\cdots U|\psi _0\rangle$, $n$ times product of $U$, is called a quantum random walk with initial state $|\psi_0\rangle$. For convenience, we will use the short notation $|x^i\rangle=|x\rangle \otimes |i\rangle$.

The probability that at time step $n$, the quantum random walk is observed at state $|x^i\rangle $ is defined by 

$$p_t(x^ i)=|\langle x^i|\psi_t\rangle|^2,$$
and  the probability that at time  $t$, the quantum random walk  is observed at state $|x\rangle $ is defined by 
\begin{eqnarray}
p_t(x)=\sum_{i=1}^2p_t(x^ i)=\sum_{i=1}^2 |\langle x^i|\psi_t\rangle|^2.
\end{eqnarray}

If the unitary operator does not depend on the position and time, the quantum walk is called homogeneous quantum walk.

%\begin{example}\label{eg:flip-flop hadamard}
%	Let $S=S_f$ and for all $x$,
%	$$
%	C=
%	\left[
%	\begin{array}{ c c  }
%	\frac{1}{\sqrt 2} &\frac{1}{\sqrt 2} \\
%	\frac{1}{\sqrt 2}&-\frac{1}{\sqrt 2}
%	\end{array} \right]
%	$$
%	Then the quantum random walk is called the flip-flop Hadamard quantum random walk.
%\end{example}

\begin{example}\label{eg:standard  hadamard}
	Let $S$ be the standard shift operator and
	$$
	C=
	\left[
	\begin{array}{ c c  }
	\frac{1}{\sqrt 2} &\frac{1}{\sqrt 2} \\
	\frac{1}{\sqrt 2}&-\frac{1}{\sqrt 2}
	\end{array} \right]
	$$
	Then the quantum random walk is called the standard Hadamard quantum random walk.
\end{example}

\subsection{Decoherent walks}\label{DecoWalk}
In physics, the quantum decoherence, in other words, lost of quantum coherence, is caused by environmental interactions. Mathematically, in quantum random walks, is defined by measurements. 
\begin{definition}
	A set of operators $\{A_n\}$ on $H$ is called a measurement if it satisfies
	$$\sum_n {A_n}^*A_n=I$$
\end{definition}
Throughout this thesis, we also assume that the measurement is unital, i.e., it satisfies
$$\sum_n A_nA^*_n=I$$

Suppose before each unitary transformation, a measurement is performed. 
After the measurement, a density operator $\rho$ on $H$ is transformed by
$$\rho\rightarrow\rho^{\prime}=\sum_n A_n\rho A^*_n.$$

Then after one step of the evolution and then under decoherence, the density operator becomes
\begin{equation}
\rho^{\prime}=\sum_nA_nU\rho U ^* A^*_n
\end{equation}
Then the decoherent quantum random walk is defined as follows

\begin{definition}
	Let $\{A_n\}$ a measurement on the Hilbert space $H=H_{p}\otimes H_c$. Suppose we start from the state $|0\rangle\otimes |\Phi_0\rangle$, with $\Phi_0\in \{1,2\}$,
	then the initial state is given by the density operator $\rho_0=|0\rangle \langle 0| \otimes |\Phi_0\rangle \langle \Phi_0|$, and the decoherent quantum walk $\{\rho_t\}_{t\in \mathbb{N}}$ with decoherence $\{A_n\}$ is defined by the following recursive relation:
	$$\rho_1=\sum_nA_nU\rho_0 U ^* A^*_n$$
	$$\rho_t=\sum_nA_nU\rho_{t-1} U ^* A^*_n$$
	
\end{definition}

We can immediately deduce from the definition and obtain that for all $t=1,2,...$  $$\rho_0=|0\rangle \langle 0| \otimes |\Phi_0\rangle \langle \Phi_0|=(|0\rangle \otimes |\Phi_0\rangle )( \langle 0| \otimes \langle \Phi_0|),$$
$$\rho_t=\sum_{n_1,\ldots,n_t}A_{n_t}U\cdots A_{n_1}U (|0\rangle \otimes |\Phi_0\rangle )( \langle 0| \otimes \langle \Phi_0|) U^*A^*_{n_1}\cdots U^*A^*_{n_t} \label{eq:rhot}$$

If we define the superoperator  $\mathcal{L}$ to be an operator which  maps $L(H)$ to $L(H)$ such that
\begin{equation}
\mathcal{L}B\equiv\sum_{n}A_{n} UB U^*A^*_{n}, \; \forall B\in L(H),
\label{eq:Lkkprime}
\end{equation}
then
\begin{equation}
\rho_t= \mathcal{L}^t\rho_0.
\end{equation}

For decoherent quantum random walk with decoherence $\{A_n\}$, the probability of reaching a point $x$ at time $t$ is defined by 
\begin{eqnarray}
p_d(x, t)&=&Tr[(|x\rangle\langle x|\otimes I_c)\rho_t]\nonumber\\
&=&Tr[ (|x\rangle\langle x|\otimes I_c)\mathcal{L}^t\rho_0],\label{eq:prob t}
\end{eqnarray}
where $Tr(\cdot)$ denotes the trace operator. 

The following examples show different decohorent quantum walks with different types of measurements $\{A_n\}_{n\in \mathbb{N}}$
\begin{example}\label{eg:total decoherent quantum random walks}
	Let $0\leq p \leq 1$, $A_0=\sqrt{1-p}I_H$, and $A_{xi}=\sqrt{p}(|x\rangle \otimes |i\rangle )(\langle x | \otimes \langle i|)$. Then $\{A_0, A_{xi}; x \in Z^1, i=1, 2\}$ is a measurement on $H$. When $0< p \leq 1$, $\rho_t$ is called a totally (coin-space) decoherent quantum random walk.
\end{example}

\begin{example}
	Let $0\leq p \leq 1$, $A_{\emptyset}=\sqrt{1-p}I_H$, and $A_{x}=\sqrt{p}|x\rangle \langle x | \otimes I_c$. Then $\{A_{\emptyset}, A_{x}; x \in Z^1\}$ is a measurement on $H$, and when $0<p\leq 1$, $\rho_t$ is called a position space decoherent quantum random walk.
\end{example}

\begin{example}\label{CoinDecoh}
	Let $0\leq p \leq 1$, $A_0=\sqrt{1-p}I_H$, and $A_{i}=\sqrt{p}I_p \otimes |i\rangle \langle i | $. Then $\{A_0, A_{i}; i=1, 2\}$ is a measurement on $H$, and when $0<p\leq 1$, $\rho_t$ is called a coin space decoherent quantum random walk.
\end{example}

Considering the general measurement $\{A_n\}$ in Example \ref{CoinDecoh} and the homogeneous unitary operator $C$, one of the recent and important achievements is that in \cite{YangConvexNormal}, S. Fan, Z. Feng, S. Xiong and W. Yang proved that under some conditions of the superoperator $\mathcal{L}$, the rescaled probability mass function on $\displaystyle\frac{\mathbb{Z}}{\sqrt{t}}$ by
$$ \hat{p}(x,t)\equiv p_d(\sqrt{t}x,t),\ x\in \frac{\mathbb{Z}}{\sqrt{t}}$$
converges in distribution to a continuous convex combination of normal distributions.

Interpreting Example \ref{eg:total decoherent quantum random walks} in physics, we perform a measurement at $|x\rangle\langle x | \otimes |i\rangle  \langle i|$ with probability $p$ and no measurement with probability $q=1-p$ at each time step $n$. If at the measurement step, the outcome is $|x\rangle \otimes |i\rangle$, then the system is reset to $|x\rangle \otimes |i\rangle$. On the other hand, we can also easily observe that if $p=1$, measurement at each time step with probability $1$, $\rho_n$ will represent the classical Markov chain or random walk. Intuitively, $\rho_n$ will be the pure quantum Markov chain or quantum walk if $p=0$.

We will mainly discuss the large scale behavior of quantum walks with the time-inhomogenuous unitary operator and the coin space decoherence measurement from Example \ref{eg:total decoherent quantum random walks}.

\section{Time-Inhomogeneous Quantum Random Walk}\label{QRW}
Percolation as a mathematical theory was introduced by Broadbent and Hammersley \cite{BroadbentPerco}, and it is applied to model probabilities which are affected by different environment. For instance space-inhomogeneous random walk on $\mathbb{Z}$, the probability of each step depends on the state position in $\mathbb{Z}$. However, instead of the space, the probability can only depends on the time step. Let's define the inhomogeneuous quantum analogue of the random walk on the infinite discrete space $\mathbb{Z}$. We let $H=H_p\otimes H_c$ and for $n=1, 2, ....$, 
$\hat{F}_n: H \to H$  be a inhomogeneous unitary transformation on $H$ defined by 
\begin{eqnarray}
\hat{F}_n=\sum_{x \in Z} |x\rangle \langle x|\otimes C_{n, x},
\end{eqnarray}

where $C_{n,x}:H_c \to H_c$ are unitary operators which depend on the time $n$ and the position $x$.

Generalizing from Definition \ref{homoQRW}, the inhomogeneous evolution operator at time $n$ of quantum random walk is given by
\begin{eqnarray}
\hat{U}_n=S\hat{F}_n,
\end{eqnarray}
where $S=S_o$ for a standard quantum random walk, and  $S=S_f$ for the flip-flop quantum random walk. 

Similarly to homogeneous quantum walk defined in previous section. Let $|\psi _0\rangle  \in H$. Then $|\psi _t\rangle =\hat{U}_t...\hat{U}_2\hat{U}_1| \psi _0\rangle$ is called an inhomogeneous quantum random walk with initial state $|\psi_0\rangle$. The probability that at time  $t$, the quantum random walk  is observed at state $|x^i\rangle $ is defined by 
\begin{eqnarray}
\hat{p}_t(x^ i)=|\langle x^i|\psi_t\rangle|^2,
\end{eqnarray}
and the probability that at time  $t$, the quantum random walk  is observed at state $|x\rangle $ is defined by 
\begin{eqnarray}
\hat{p}_t(x)=\sum_i |\langle x^i|\psi_t\rangle|^2.
\end{eqnarray}

For decoherent time-inhomogeneous quantum random walks, we 
suppose the quantum walk starts at the state $|0\rangle\otimes |\Phi_0\rangle$,
then, the initial state is given by the density operator
\begin{equation}
\hat{\rho}_0=|0\rangle \langle 0| \otimes |\Phi_0\rangle \langle \Phi_0|=(|0\rangle \otimes |\Phi_0\rangle )( \langle 0| \otimes \langle \Phi_0|).
\end{equation}
After $t$ steps, with the decoherence measurement $\{A_n\}_{n\in\mathbb{N}}$, the state evolves to
\begin{eqnarray}
\hat{\rho}_t=\sum_{n_1,\ldots,n_t}A_{n_t}\hat{U}_t\cdots A_{n_1}\hat{U}_1 (|0\rangle \otimes |\Phi_0\rangle )( \langle 0| \otimes \langle \Phi_0|) \hat{U}_1^*A^*_{n_1}\cdots \hat{U}_t^*A^*_{n_t}\\ \label{eq:inho rhot}
=\sum_{n}A_{n}\hat{U}_t\hat{\rho}_{t-1} \hat{U}_t^*A^*_{n}
\end{eqnarray}

The probability that at time  $t$, the decoherent inhomogeneous quantum random walk   is observed at state $|x^i\rangle $ is defined by 
\begin{eqnarray}
\hat{p}_t(x^ i)=Tr([|x^i\rangle  \langle x^i|] \hat{\rho}_t)
\end{eqnarray}
and  the probability that at time  $t$, the quantum random walk  is observed at state $|x\rangle $ is defined by 
\begin{eqnarray}
\hat{p}_t(x)=\sum_{i}p_t(x^ i)=Tr([|x\rangle  \langle x|\otimes I_c] \hat{\rho}_t)\label{eq:inho rhot prob}
\end{eqnarray}

If the unitary operator depends on the position or the time, the quantum walk is called inhomogeneous quantum walk, here we have some examples of position-inhomogenuous quantum walk and time-inhomogenuous quantum walks,

%\begin{example}\label{eg:flip-flop linear drift}
%	Let $S=S_f$, and $C_{n,x}=C_x$ depends on the position such that
%	$$
%	C_x=
%	\left[
%	\begin{array}{ c c  }
%	(\frac{1}{1+|x|})^{1/2} &(1-\frac{1}{1+|x|})^{1/2} \\
%	(1-\frac{1}{1+|x|})^{1/2}&-(\frac{1}{1+|x|})^{1/2}
%	\end{array} \right]
%	$$
%	Then the quantum random walk is called the flip-flop quantum random walk with linear drift. 
%\end{example}

%\begin{example}\label{eg:standard linear drift}
%	Let $S=S_o$, and $C_{n,x}=C_x$ depends on the position such that
%	$$
%	C_x=
%	\left[
%	\begin{array}{ c c  }
%	(\frac{1}{1+|x|})^{1/2} &(1-\frac{1}{1+|x|})^{1/2} \\
%	(1-\frac{1}{1+|x|})^{1/2}&-(\frac{1}{1+|x|})^{1/2}
%	\end{array} \right]
%	$$
%	Then the quantum random walk is called the standard  quantum random walk with linear drift. 
%\end{example}

\begin{example}\label{timeInhomoEx}
	Let $C_{n,x}=C_n$ depends on the time step such that
	$$
	C_n=\begin{bmatrix} 
	\sqrt{1-\frac{\lambda}{n^{\zeta}}} & \sqrt{\frac{\lambda}{n^{\zeta}}} \\
	\sqrt{\frac{\lambda}{n^{\zeta}}} & -\sqrt{1-\frac{\lambda}{n^{\zeta}}} 
	\end{bmatrix} 
	$$
	where $\lambda$ and $\zeta$ are non negative real numbers.
\end{example}

\subsection{Time-inhomogenuous quantum walk and its path integral expression}
We now concentrate on analyzing the decoherent time-inhomogeneous quantum walk defined with unitary operators in Example \ref{timeInhomoEx}, and the total decoherence measurement defined in Example \ref{eg:total decoherent quantum random walks}. We have with the initial density operator $\rho_0=|0\rangle \langle 0| \otimes |\Phi_0\rangle \langle \Phi_0|$, and the decoherence measurement 
$$\{ A_{xi}; x \in Z^1, i=1, 2\}\cup\{A_0\}$$ where for $p\in [0,1]$, $x\in \mathbb{Z}$, and $i=1,2$ 
$$A_{xi}=\sqrt{p}\cdot|x\rangle\langle x | \otimes |i\rangle  \langle i|,$$ and 
$$A_0=\sqrt{1-p}\cdot I_p\otimes I_c.$$ 
Then our time inhomogeneous quantum walk with total decoherence measurement with decoherence parameter $p$ is defined as 
\begin{eqnarray}\label{InhomoWalkwithlamdazeta}
\hat{\rho}_t=A_{0}\hat{U}_t\hat{\rho}_{t-1} \hat{U}_t^*A^*_{0}+\sum_{x,i}A_{x,i}\hat{U}_t\hat{\rho}_{t-1} {\hat{U}_t}^*A^*_{x,i} \label{eq:inhodechowalk}
\end{eqnarray}
with the time-inhomogeneous unitary operators
\begin{eqnarray}\label{InhomoOperator}
C_n=\begin{bmatrix} 
\sqrt{1-\frac{\lambda}{n^{\zeta}}} & \sqrt{\frac{\lambda}{n^{\zeta}}} \\
\sqrt{\frac{\lambda}{n^{\zeta}}} & -\sqrt{1-\frac{\lambda}{n^{\zeta}}} 
\end{bmatrix}
\end{eqnarray}
where $\lambda,\zeta>0$.

Let $T_1, T_2...$ geometric random variables with probability $p$ with $\sigma_1=T_1$,..., $\sigma_n=T_1+...+T_n$, and $\rho_0=|0\rangle\langle 0| \otimes|i_0\rangle \langle i_0|$, we have that the probability after $t$ steps at the position $x$
$$\hat{p}_t(x) = \sum_{j=1}^{2}Tr\Big(|x\rangle\langle x|\otimes |j\rangle \langle j|\hat{\rho}_t\Big)$$

Recall the $Q_{\sigma_{n}}$'s defined by C. Chou and W. Yang in \cite{ChouQMarkov} using the time-inhomogeneous operators above in (\ref{InhomoOperator})
\begin{eqnarray}
Q_{\sigma_{n-1}}(i,j):=E\Big[\big|\langle j| C_{\sigma_{n-1}+T_n}\cdots C_{\sigma_{n-1}+1} |i\rangle\big|^2\Big],
\end{eqnarray}
and, for discrete infinite space quantum walks, we generalize the idea to the following definition,

\begin{definition}
	Let $\hat{Q}_{\sigma_i\sigma_{i+1}}(x,y,i,j)$ the probability from  $x$ to the state $y$ on the position space, and $i$ to $j$ on the coin space during the time $\sigma_i$ to $\sigma_{i+1}$, and $\hat{W}(x,y,i,j)$ the probability from  $x$ to the state $y$ on the position space, and $i$ to $j$ on the coin space during the time $\sigma_n$ to $t$, which is
	$$\hat{Q}_{\sigma_k,\sigma_{k+1}}(x,y,i,j)=|\langle y,j|\hat{U}_{\sigma_{k+1}}\cdots \hat{U}_{\sigma_k+1}|x,i\rangle|^2$$ 
	$$\hat{W}_{\sigma_{N_t}, t}(x,y,i,j)=|\langle y,j|\hat{U}_{t}\cdots \hat{U}_{\sigma_{N_t}}|x,i\rangle|^2$$
	where $x,y\in \mathbb{Z}$, and $i,j=1,2$
	
\end{definition}

Therefore, we have the probability at $x$ after $t$ time steps using the path integral expression, by coin-space decoherence
$$\hat{p}_t(x)=\sum_{j=1}^{2}E\Big[\sum_{x_1,...,x_{\sigma_{N_t}}\in \mathbb{Z}}\sum_{i_1,...i_{N_t}\in \{1,2\}}\hat{Q}_{\sigma_{0}\sigma_{1}}(0,i_0,x_1,i_1)\cdots$$
$$\cdots\hat{Q}_{\sigma_{N_t-1}\sigma_{N_t}}(x_{\sigma_{N_t}-1},i_{\sigma_{N_t}-1},x_{N_t},i_{N_t})\hat{W}_{\sigma_{N_t},t}(x_{\sigma_{N_t}},i_{\sigma_{N_t}},x,j)\Big]$$
Note that using the translation invariant property $$\hat{Q}_{\sigma_{i}\sigma_{i+1}}(x,i,y,j)=\hat{Q}_{\sigma_{i}\sigma_{i+1}}(0,i,y-x,j),$$
$$\hat{W}_{\sigma_{N_t},t}(x,i,y,j)=\hat{W}_{\sigma_{N_t},t}(0,i,y-x,j)$$
Let's just denote 
$$\hat{Q}_{\sigma_{i}\sigma_{i+1}}(x,i,j):=\hat{Q}_{\sigma_{i}\sigma_{i+1}}(0,i,x,j),$$ and
$$\hat{W}_{\sigma_{N_t},t}(x,i,j):=\hat{W}_{\sigma_{N_t},t}(0,i,x,j),$$
and define for $i,j \in \{1,2\}$
$$R_{\sigma_{k}\sigma_{k+1}}(i,j):=\sum_{x\in \mathbb{Z}}\hat{Q }_{\sigma_{k}\sigma_{k+1}}(x,i,j)$$
$$\tilde{R}_{\sigma_{N_t},t}(i,j):=\sum_{x\in \mathbb{Z}}\hat{W }_{\sigma_{N_t},t}(x,i,j)$$
And, we obtain the path integral expression for $\hat{p}_t(x)$,
$$\hat{p}_t(x)=\sum_{j=1}^{2}E\Big[\sum_{x_1,...,x_{k}}\sum_{i_1,...i_{k}}\frac{\hat{Q}_{\sigma_{0}\sigma_{1}}(x_1,i_0,i_1)}{R_{\sigma_{0}\sigma_{1}}(i,i_1)}\cdots\frac{\hat{Q}_{\sigma_{N_t-1}\sigma_{N_t}}(x_{k}-x_{k-1},i_{k-1},i_{k})}{R_{\sigma_{N_t-1}\sigma_{N_t}}(i_{k-1},i_{k})}\cdot$$
$$\cdot\frac{\hat{W}_{\sigma_{N_t}t}(x-x_{k},i_{k},j)}{\tilde{R}_{\sigma_{N_t}t}(i_{k},j)}\cdot R_{\sigma_{0}\sigma_{1}}(i_0,i_1)\cdots R_{\sigma_{N_t-1},\sigma_{N_t}}(i_{\sigma_{N_t}-1},i_{\sigma_{N_t}})\tilde{R}_{\sigma_{N_t}t}(i_{k},j)\Big]$$

\subsection{Representation theorem}
Now, suppose that  $I_k$ is the Markov chain defined by $I_0=i_0$ with the property that $P(I_{k+1}=j|I_k=i)=R_{\sigma_{k}\sigma_{k+1}}(i,j)$, and $\tilde{I}_t$ such that $P(I_{t}=j|I_{N_t}=i)=\tilde{R}_{\sigma_{N_t}t}(i,j)$. 
Also let $\mu_{\sigma_{k}\sigma_{k+1}}(\cdot,i,j)$ and $\tilde{\mu}_{\sigma_{N_t}t}(\cdot,i,j)$ be probability distributions on $\mathbb{Z}$ for $i,j=1,2$ defined as
$$\mu_{\sigma_{k}\sigma_{k+1}}(x,i,j) :=  \frac{\hat{Q}_{\sigma_k,\sigma_{k+1}}(x,i,j)}{\sum_{x\in \mathbb{Z} }\hat{Q}_{\sigma_k,\sigma_{k+1}}(x,i,j)},$$
and
$$\tilde{\mu}_{\sigma_{N_t}t}(x,i,j)=\frac{\hat{W}_{\sigma_{N_t},t}(x,i,j)}{\sum_{x\in\mathbb{Z}}\hat{W}_{\sigma_{N_t},t}(x,i,j)},$$
Let $Y_{\sigma_{k},\sigma_{k+1}}(i,j)$  independent random variables with distributions $\mu_{\sigma_{k}\sigma_{k+1}}(\cdot,i,j)$, and $\tilde{Y}_{\sigma_{N_t},t}(i,j)$ random variable with distribution $\tilde{\mu}_{\sigma_{N_t}t}(\cdot,i,j)$. We have

$$\hat{p}_t(x)=\sum_{j=1}^{2}E^{\sigma}E^{Y}\Big(\mathds{1}_x\big[Y_{\sigma_{0}\sigma_{1}}(i_0,i_1)+Y_{\sigma_{1}\sigma_{2}}(i_1,i_2)+\cdots + Y_{\sigma_{N_t-1}\sigma_{N_t}}(i_{k-1},i_k)+$$
$$+\tilde{Y}_{\sigma_{N_t}t}(i_{k},j)\big]\cdot\big[R_{\sigma_{0}\sigma_{1}}(i_0,i_1)\cdots R_{\sigma_{N_t-1}\sigma_{N_t}}(i_{k-1},i_k)\cdot\tilde{R}_{\sigma_{N}t}(i_{k},j)\big]\Big)$$
$$=E^\sigma E^IE^Y\Big(\mathds{1}_x\Big[Y_{\sigma_{0}\sigma_{1}}(i_0,I_1)+Y_{\sigma_{1}\sigma_{2}}(I_1,I_2)+\cdots+Y_{\sigma_{N_t-1}\sigma_{N_t}}(I_{N_t-1},I_{N_t})+$$
$$+\tilde{Y}_{\sigma_{N_t}t}(I_{N_t},I_t)\Big]\Big)$$

We have proved the following representation theorem,

\begin{theorem}[$\sigma$-$I$-$Y$ representation theorem]\label{RepreQuant}
	Let the initial state be $|0\rangle\otimes|i_0\rangle$, then the probability of the time-inhomogeneous quantum walk defined in (\ref{InhomoWalkwithlamdazeta}) and (\ref{InhomoOperator}) is found in $x$ is 
	$$\hat{p}_t(x)=E^\sigma E^IE^Y\Big(\mathds{1}_x\Big[Y_{\sigma_{0}\sigma_{1}}(i_0,I_1)+Y_{\sigma_{1}\sigma_{2}}(I_1,I_2)+\cdots+Y_{\sigma_{N_t-1}\sigma_{N_t}}(I_{N_t-1},I_{N_t})+$$
	$$+\tilde{Y}_{\sigma_{N_t}t}(I_{N_t},I_t)\Big]\Big)$$
\end{theorem}

\begin{remark}
	Theorem \ref{RepreQuant} not only gives us a new formula about calculating the probability of a quantum random walk to be found in $x$ at time $t$, but also a better visualization of it using path integral expression connecting quantum probability and classical analytic probability.
	
\end{remark}

\section{Applications and examples}\label{ApplicationRW}
Theorem \ref{RepreQuant} directly implies the following Monte Carlo simulation algorithms to estimate the probability at $x$ and the distribution of the decoherent quantum walk at time $t$. Following the proof of Theorem \ref{RepreQuant}, and using the same notations, we have
\begin{enumerate}[label={\bfseries Step \arabic*:}]
	\item Fix $t$, generate $T_1,T_2,...,T_n,T_{n+1}$ iid with geometric distribution with probability $p$, and let $\sigma_0=0, \sigma_1=T_1,...,\sigma_{n+1}=T_1+\cdots +T_{n+1}$, and suppose that $\sigma_n<t<\sigma_{n+1}$.
	\item Let $ \hat{Q}_{\sigma_k,\sigma_{k+1}}(x,i,j)=|\langle x,j|\hat{U}_{\sigma_{k+1}}\cdots \hat{U}_{\sigma_k+1}|0,i\rangle|^2$, and $ \hat{W}_{\sigma_{n},t}(x,i,j)=|\langle x,j|\hat{U}_{t}\cdots \\ \hat{U}_{\sigma_{n}}|0,i\rangle|^2$ where $x\in \mathbb{Z}$, and $i,j=1,2$.
	%\item Consider $\displaystyle \mu_{\sigma_{k}\sigma_{k+1}}(x,i,j) =  \frac{\hat{Q}_{\sigma_k,\sigma_{k+1}}(x,i,j)}{\sum_{x\in \mathbb{Z} }\hat{Q}_{\sigma_k,\sigma_{k+1}}(x,i,j)}$.
	%\item Consider $\displaystyle \tilde{\mu}_{\sigma_{N_t},t}(x,i,j) =  \frac{\hat{W}_{\sigma_{N_t},t}(x,i,j)}{\sum_{x\in \mathbb{Z} }\hat{W}_{\sigma_{N_t},t}(x,i,j)}$.
	\item Using the $\hat{Q}_{\sigma_k,\sigma_{k+1}}$ and $ \hat{W}_{\sigma_{n},t}$ from the previous step to generate $Y_{k+1}$, independent random variables with distribution $\mu_{\sigma_{k}\sigma_{k+1}}(\cdot,i,j)$, and $\tilde{Y}_t$ random variable with distribution $\tilde{\mu}_{\sigma_{n},t}(.,i,j)$.
	\item Fix $i$, and $j$, generate $Z_t(i,j) = Y_1(i,j)+Y_2(i,j)+\cdots+Y_{n}(i,j)+\tilde{Y}_t(i,j)$
	\item Generate different samples of the Markov chain $I_0=i_0$, $\{I_k\}_{k=0}^n$, and $I_t$ with the transition $P(I_{k+1}=j|I_k=i)=R_{\sigma_{k}\sigma_{k+1}}(i,j)$, and $P(I_{t}=j|I_{n}=i)=\tilde{R}_{\sigma_{n},t}(i,j)$ for each Markov chain $I_t$ generate sample for $Z_t(i_0,I_t)$.
	\item Repeat the procedure with different samples of $\{\sigma_n\}$, and take the average over  $Y$, $I$, and $\sigma$, then we obtain the probability for each $x\in\mathbb{Z}$, and the distribution of the decoherent quantum walk at time $t$.
	
\end{enumerate}

We run the simulation using the algorithm with different values of $\lambda$, $\zeta$ and $p$ in Python, and following examples illustrate simulated scaling limits approximate the theoretical results proven with $t=500$ and number of samples $500$, $500$ and $2000$ for $Y, \sigma$ and  $I$ respectively.

\subsection{Approximation of classical probability distribution densities}

We note that if $p=1$, the probability to make the measurement at each step is 1 which means that the decoherent quantum walk becomes a classical probability random walk with time-inhomogeneous transition matrix
$$
C_n=\begin{bmatrix} 
1-\frac{\lambda}{n^{\zeta}} & \frac{\lambda}{n^{\zeta}} \\
\frac{\lambda}{n^{\zeta}} & 1-\frac{\lambda}{n^{\zeta}} 
\end{bmatrix}, 
$$
and J. Englander and S. Volkov in \cite{EnglanderClassic} proved that if $\zeta=1$, the scaling limit 
$$\hat{p}(x,t)\equiv p_d(tx,t),\ x\in \frac{\mathbb{Z}}{t}$$
converges to the symmetric Beta distribution, {\bfseries Beta($\lambda$,$\lambda$)}, in $[-1,1]$. In particular, $\hat{p}(x,t)$ converges to arcsine law, uniform law and semicircle law when $\lambda=\frac{1}{2}, 1, \frac{3}{2}$ respectively.

\begin{figure}[h!]
	\centering
	\includegraphics[width=0.6\textwidth]{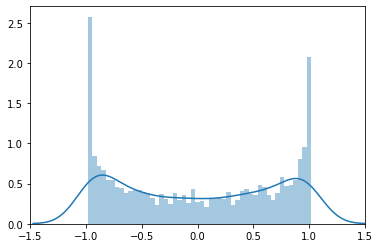}
	\caption{Approximated arcsine law for $\zeta=1$, $\lambda=\frac{1}{2}$ and $p=1$}
	\label{fig:RW1}
\end{figure}

Therefore in our model, with $p=1$ and respective parameters above, our algorithm generates the approximations of these distributions. By taking large samplings numbers and time scales, the generated approximated distributions will converge to the theoretical distributions. The shaped area restricted in the interval $[-1,1]$ in Figure \ref{fig:RW1}
%, Figure \ref{fig:RW2} and Figure \ref{fig:RW3}
shows the obtained approximated arcsine
%, uniform, and 
law as the simulation result.

\subsection{Approximation of decoherent Hadamard walk}

We can observe that if $\zeta=0$, the model becomes a decoherent homogeneous quantum walk. For instance, by taking $\zeta=0$ and $\lambda = \frac{1}{2}$, the unitary operators will be 

$$C_n=\begin{bmatrix} 
\sqrt{\frac{1}{2}} & \sqrt{\frac{1}{2}} \\
\sqrt{\frac{1}{2}} & -\sqrt{\frac{1}{2}} 
\end{bmatrix} $$
for all $n$, and we have the decoherent Hadamard walk with both coin and position spaces measurement. K. Zhang in \cite{ZhangQuantumLimit} proved that the limiting distribution of the rescaled probability mass function on $\displaystyle\frac{\mathbb{Z}}{\sqrt{t}}$ by
$$ \hat{p}(x,t)\equiv p_d(\sqrt{t}x,t),\ x\in \frac{\mathbb{Z}}{\sqrt{t}},$$
is Gaussian with mean $\mu=0$, and variance $\displaystyle\sigma^2=\frac{p+2\sqrt{1+q^2}-2}{p}$ where $q=1-p$. 

\begin{figure}[h!]
	\centering
	\includegraphics[width=0.6\textwidth]{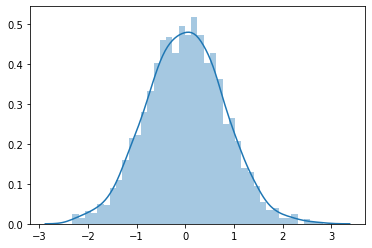}
	\caption{Approximated density for $\zeta=0$, $\lambda=\frac{1}{2}$ and $p=\frac{1}{2}$}
	\label{fig:RW4}
\end{figure}

Figure \ref{fig:RW4} shows the obtained the respective approximated normal distribution with $\zeta=0$, $\lambda=\frac{1}{2}$ and $p=\frac{1}{2}$ generated by our algorithm.

\subsection{Estimating limiting distributions when $p<1$}

Physical intuitions tell us that whenever the pure quantum system starts being interacted by the environment, the system will become classical phenomenons in long term, in other words, the measurements can make the pure quantum system approximating the classical results. Mathematically, if the decoherent parameter $p$ is greater than $0$, the long time scaling limits should be similar to the pure classical case when $p=1$, analogue results for decoherent analogue quantum walk were proved in \cite{YangConvexNormal} and \cite{ZhangQuantumLimit}.

First, we consider the case $\zeta=1$. However, unlike the decoherent Hadamand walk case, for $\zeta=1$, arcsine, uniform, and semicircle law in the interval $[-1,1]$ have no parameters, which we can deduce that it will be difficult to find the explicit limiting distribution for $p<1$. Thus, finding the right scaling parameter will be crucial for these cases, but we observe that if the scaling limit is $n^\alpha$ where $\alpha\neq1$, the densities will spread out to either infinity or accumulate to only one point. Therefore, we estimate the decoherent densities for $0<p<1$ with $\lambda=\frac{3}{2}$, with the scaling exponent $\alpha=1$.

\begin{figure}[h!]%
	\centering
	\subfigure[$p=0.3$]{%
		\label{fig:RW8a}%
		\includegraphics[height=1.75in]{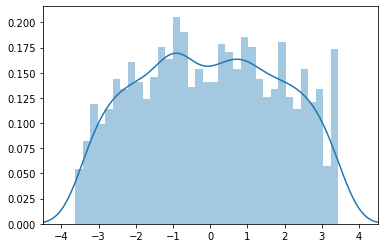}}%
	\hfill
	\subfigure[$p=0.7$]{%
		\label{fig:RW8b}%
		\includegraphics[height=1.75in]{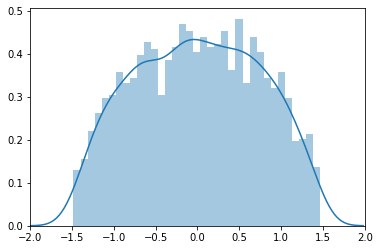}}%
	\caption{Estimated partially decoherent densities for $\zeta=1$ and $\lambda=\frac{3}{2}$}
	\label{fig:RW8}
\end{figure}
%\textcolor{blue}{Here, I reduced the number of figures from all arcsine, uniform and semicircle to one, semicircle}

The shaped areas in Figure
 %\ref{fig:RW6}, \ref{fig:RW7}, and
 \ref{fig:RW8} illustrats the transitions of $\zeta=1$ and $\lambda= \frac{3}{2}$ (semicircle law case) when $p$ increases. Since it is not parameterizable, we observe the transitions and compare with the classical = semicircle density. Here, it is clear that they have the classical distribution shapes, and when the decoherence parameter $p$ increases, the estimated densities approximate to the classical distributions.

J. Englander and S. Volkov(\cite{EnglanderClassic}) also proved that if $0<\zeta<1$, the scaling limit
$$\hat{p}(x,t)\equiv p_d(t^{\frac{(1+\zeta)}{2}}x,t),\ x\in \frac{\mathbb{Z}}{t^{\frac{(1+\zeta)}{2}}}$$
converges to {\bfseries Normal($0,\sigma^2$)} where $\displaystyle\sigma=\frac{1}{\sqrt{\lambda(1-\zeta)}}$. Which means that the limiting distribution is Gaussian with parameter $\sigma^2$ when $p=1$. Like the homogeneous case, we expect now that the the variance of the limiting distribution may also depend on the decoherence parameter $p$. Even though there is no rigorous proofs about their explicit limiting distributions, we estimate the limiting distributions through our algorithm.

\begin{figure}[h!]%
	\centering
	\subfigure[$p=0.3$]{%
		\label{fig:RW9a}%
		\includegraphics[width=2.5in,height=1.3in]{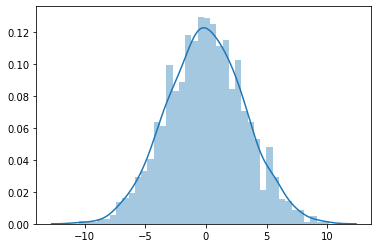}}%
	\hfill
	\subfigure[$p=0.7$]{%
		\label{fig:RW9b}%
		\includegraphics[width=2.5in,height=1.3in]{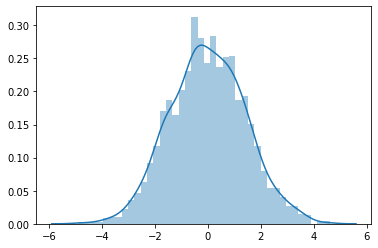}}%
	\hfill
	\subfigure[$p=1$]{%
		\label{fig:RW9c}%
		\includegraphics[width=2.5in,height=1.3in]{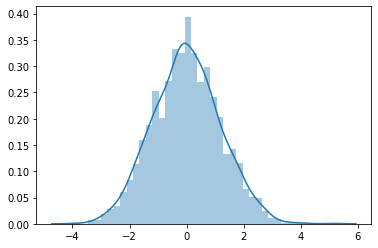}}%
	\caption{Estimated partially decoherent densities for $\zeta=\frac{1}{5}$ and $\lambda=\frac{1}{2}$}
	\label{fig:RW9}
\end{figure}

Figure \ref{fig:RW9}
%, \ref{fig:RW10}, and \ref{fig:RW11} 
shows the simulation results for estimated densities for $\zeta=\frac{1}{5}$ with different $p$. These figures not only illustrate the transitions when $p$ increases, but also demonstrate the normality of the densities. We observe that the estimated limiting distributions are Gaussian, and the variances are greater when $p$ is small.
%Moreover, from Figure \ref{fig:RW11}, we note that the convergence is slower. 
%Therefore we expect that the limiting distributions are normal, and the variances depend on the parameter $p$ such that if $p$ decreases, the variances increase.

%Last case we consider is when $\lambda>1$, and Figure \ref{fig:RW12} illustrate the estimated densities for $p=0.3$ and $p=0.7$.

%\begin{figure}[h!]%
%	\subfigure[$p=0.3$]{%
%		\label{fig:RW12a}%
%		\includegraphics[height=1.8in]{QRWz15lam05p07t500numSam500numSampI2000}}%
%	\hfill
%	\subfigure[$p=7$]{%
%		\label{fig:RW12b}%
%		\includegraphics[height=1.8in]{QRWz15lam05p10t500numSam500numSampI2000}}%
%	\caption{Estimated partially decoherent densities for $\zeta=\frac{3}{2}$ and $\lambda=\frac{1}{2}$}
%	\label{fig:RW12}
%\end{figure}

Note that the applications of Theorem \ref{RepreQuant} give us not only a analytic visualization of decoherent quantum walks in general, but also an approach to approximate classical distributions through quantum algorithms which could be useful in the future when quantum computers are fully developed.

\section{Quantizing classical distributions}\label{QCD}
It is time to consider the time-inhomogeneous pure quantum walk. In this case,  the probability to make a measurement at each time step is $p=0$, the system maintains in the pure quantum environment. On the other hand, we know that for the complete space-time decoherent quantum walk $p=1$ from previous section, the limiting distributions with the appropriate scaling exponents, converge to 
\begin{itemize}
	\item {\bfseries Beta($\lambda$,$\lambda$)} law when $\zeta=1$.
	\item {\bfseries Bernoulli($\frac{1}{2}$)} law when $\zeta>1$.
\end{itemize}

As results, our model gives us an algorithm to determine quantum analogues of these two classical distributions by taking $p=0$ with respective parameters mentioned above. We call them quantized classical distributions. For instance, we obtain a quantized normal distribution by considering the Hadamard Walk, and a quantized arcsine distribution by taking $p=0$, $\zeta=1$, and $\lambda=\frac{1}{2}$.

Unlike the decoherent quantum walk, the pure quantum walk calculation can be executed easily by linear algebra and matrices operations. Therefore, the explicit distributions are obtained by simulation, and the quantum analogues of the classical distributions are visualized and analyzed numerically. 

\subsection{Quantized Beta and Bernoulli distributions when $\zeta\geq 1$}
The fact that the density of Hadamard walks, with scaling exponent $1$ with symmetric initial conditions, converges to  

$$\frac{1}{\pi(1+x)\sqrt{1-2x^2}} \ \text{ for } \  x\in(-\frac{1}{\sqrt{2}},\frac{1}{\sqrt{2}}),$$

proven by Konno in \cite{KonnoHadamardLimit} (Quantized normal distribution) motivates us to study numerically the convergence of
\begin{eqnarray}\label{rescaledDistribution}
\hat{p}(x,t)\equiv p_d(t^{\gamma}x,t),\ x\in \frac{\mathbb{Z}}{t^{\gamma}},
\end{eqnarray}
with $p=0$ and different $\zeta$'s and $\lambda$'s, distributions of time-inhomogeneous pure quantum random walk. We will analyze statistically the scaling limits and the convergence rates $\gamma$ of Equation (\ref{rescaledDistribution}).

%\subsection{Scaling exponents for }
We first consider the case $\zeta\geq 1$, when the classical distributions do not converge to normal distributions. Using symmetric initial conditions,

$$\rho_0=\frac{1}{\sqrt{2}}\big(|0 \rangle \otimes |1\rangle + |0 \rangle \otimes |2\rangle\big),$$

%\textcolor{blue}{Here, I eliminated some figures which initially illustrated the convergence of the mass}

%Figures \ref{fig:RW13}, \ref{fig:RW14},  \ref{fig:RW15} and \ref{fig:RW16} illustrate the simulation results of the quantized arcsine, uniform, semicircle , and Bernoulli law using different scaling exponents $t^\alpha$ with $\alpha=0.7, 0.8 , 0.9, 1$ for $t=2000$.
%\begin{figure}[h!]
%	\centering
%	\includegraphics[width=0.8\textwidth]{pureQuantum/zeta10lambda05t2000}
%	\caption{Explicit density for $\zeta=1$, $\lambda=\frac{1}{2}$ using different scaling exponents $\alpha= 0.7, 0.8, 0.9, 1$}
%	\label{fig:RW13}
%\end{figure}

%\begin{figure}[h!]
%	\centering
%	\includegraphics[width=0.8\textwidth]{pureQuantum/zeta10lambda10t2000}
%	\caption{Explicit density for $\zeta=1$, $\lambda=1$ using different scaling exponents $\alpha= 0.7, 0.8, 0.9, 1$}
%	\label{fig:RW14}
%\end{figure}

%\begin{figure}[h!]
%	\centering
%	\includegraphics[width=0.8\textwidth]{pureQuantum/zeta10lambda15t2000}
%	\caption{Explicit density for $\zeta=1$, $\lambda=\frac{3}{2}$ using different scaling exponents $\alpha= 0.7, 0.8, 0.9, 1$}
%	\label{fig:RW15}
%\end{figure}

%\begin{figure}[h!]
%	\centering
%	\includegraphics[width=0.8\textwidth]{pureQuantum/zeta15lambda05t2000}
%	\caption{Explicit density for $\zeta=\frac{3}{2}$, $\lambda=\frac{1}{2}$ using different scaling exponents $\alpha= 0.7, 0.8, 0.9, 1$}
%	\label{fig:RW16}
%\end{figure}

Simulation results that the mass of the densities spreads rapidly to the end points of the interval. Because of the fact that $t^r \rightarrow \infty$ for $r>0$, the mass will spread out to infinity if $\alpha<1$, we expect that the correct scaling exponent is $1$. Moreover, we also observe that even for the scaling exponent equal to $1$, the mass will concentrate at the a neighborhood of the end points of the interval $[-1,1]$. In the following section, we analyze numerically that the mass will concentrate at the end points of the interval $[-1,1]$ with scaling exponent $\alpha=1$.

\subsection{Convergent rates to Bernoulli distribution}\label{QuantumToBernoulli}

Let's statistically analyze the convergent rates, by fixing a large $t$, suppose that $\alpha= 0.03$ and define 
$$\epsilon_t:=\min\{k\in \mathbb{N}: \sum_{i=-t}^{-t+k}p_d(i,t) \geq \frac{1-\alpha}{2}\},$$
Intuitively, since we are using symmetric initial conditions, for fix $t$, more than $97\%$ of the density in the set 
$$\{x\in\mathbb{Z}: x=- t,-t+1,..., -t+ \epsilon_t\}\cup\{x\in\mathbb{Z}: x=t-\epsilon_t,...,t-1,t\},$$
In other words, $\epsilon_t$ is the length of the neighborhood from the end points $-t$ and $t$ such that $97\%$ of the mass is concentrated.

And now we define $\alpha_t=\frac{\epsilon_t}{t}$. Note that in this case, with fixed t, the probability that you can find the rescaled quantum walk in the interval in $[-1,-1+\alpha_t]\cup[1-\alpha_t,1]$ is $1-\alpha$, which is,

$$\sum_{x\in[-1,-1+\alpha_t]}\hat{p}(x,t)\geq \frac{1-\alpha}{2} \ \text{ where } \ \  x\in \frac{\mathbb{Z}}{t}.$$

Assuming the fact that $\alpha_t$ converges to $0$ as $t\to \infty$, we fit two  nonlinear regression: exponential decay model $ce^{-rt}$ and rational decay model $ct^{-r}$ in Matlab. (See \cite{NonLinearStats} \cite{NonLinearTools} for more details about nonlinear regression and its applications)

First, suppose that 
$$\alpha_t\sim ce^{-rt},$$
taking natural logarithms both side, we obtain that
$$\ln(\alpha_t)\sim \ln(c)+(-rt).$$

Therefore, the exponential decay rate $r$ can be obtained by
$$r= \lim_{t\to \infty} \frac{\ln(\alpha_t)}{t},$$  
in other words, if $t$ is large, $r$ is approximately
$$r\sim \frac{\ln(\alpha_t)}{t}.$$
Second, if we assume the rational decay model $\alpha_t\sim ct^{-r}$, we can obtain by similar argument that the rational decay rate,
$$r= \lim_{t\to \infty} \frac{\ln(\alpha_t)}{\ln(t)},$$
and if $t$ is large , $r\sim \frac{\ln(\alpha_t)}{\ln(t)}.$

Considering $t=2000$, by taking initial vectors 
$$[c_0,r_0]=\displaystyle [1,\frac{\ln(\alpha_{2000})}{2000}], [1,\frac{\ln(\alpha_{2000})}{\ln(2000)}]$$ 
for exponential and rational models respectively, we fit both models in Matlab. As results, rational decay model has better R-squared estimate and root mean squared error. For instance, with 
%(see definitions of R-squared estimate and root mean squared and Figure \ref{fig:Nonlinear03} in Appendix \ref{nonLinearApenn}). For instance, Figures \ref{fig:Nonlinear01a} and \ref{fig:Nonlinear01b} show better results for rational decay model when
$\zeta=1$ and $\lambda=\frac{1}{2}$,
the R-squared estimate of the rational decay model is $0.987$, while the R-squared estimate is $0.915$ for the exponential decay model. The root mean error: $0.0154$ for the rational decay model and $0.0389$ for the exponential decay model.

\begin{figure}[h!]%
	\centering
	\subfigure[$\lambda=1/2$]{%
		\label{fig:RW18:a}%
		\includegraphics[width=2.8in,height=2.3in]{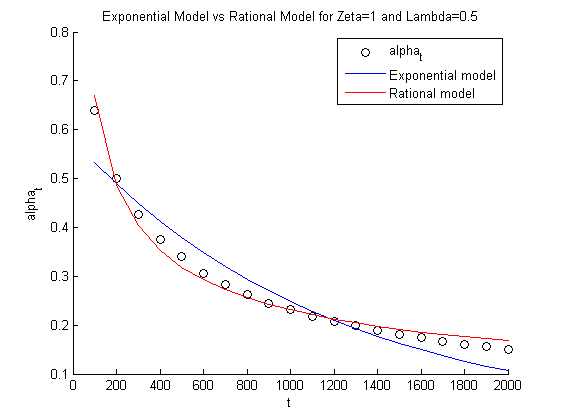}}%
	\hfill
	\subfigure[$\lambda=1$]{%
		\label{fig:RW18:b}%
		\includegraphics[width=2.8in,height=2.3in]{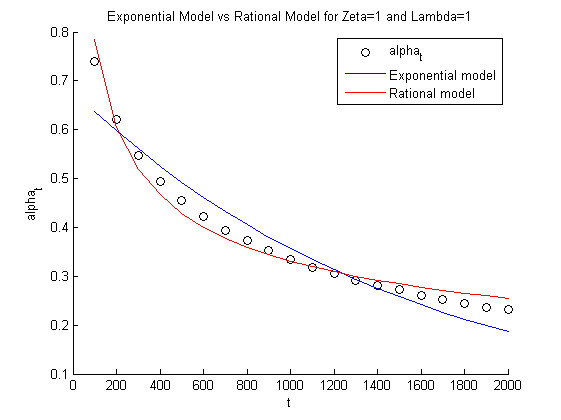}}%
	\caption{Non linear regression model comparison for $\zeta=1$ and $p=0$}
	\label{fig:RW18}
\end{figure}

On the other hand, Figure \ref{fig:RW18} shows both fitted model and $\alpha_t$ with $\lambda=\frac{1}{2}, 1$ and fixed $\zeta=1$, and we observe that the rational decay model fits better . It is clear that the functions graphed by rational model using estimated coefficients by nonlinear regression behave more similarly than the functions graphed by exponential models.

Note that how fast $\alpha_t$ converges to $0$ determines the rate of convergence to Bernoulli distribution for each $\zeta$ and $\lambda$. In order to compare numerically the rates for different values, we execute nonlinear regression for the rational model by fixing $\zeta=1$ and varying $\lambda$. then by varying $\zeta$ and fixing $\lambda=\frac{1}{2}$. Tables \ref{tab:RW01} and \ref{tab:RW02} present estimated values of the nonlinear regression for the rational model $\alpha_t\sim ct^{-r}$ we considered.

\begin{table}[h!]
	\centering
	\begin{tabular}{ |c|c|c|c|c|c|c|c|c|c|c|c| } 
		\hline
		$\lambda$ & 0.5 & 0.6 & 0.7 & 0.8  & 0.9 & 1 & 1.1 & 1.2 & 1.3 & 1.4 & 1.5 \\ 
		\hline
		c & 5.64 & 5.25 & 5.17 & 4.70  & 4.51  & 4.40 & 4.20 & 4.08 & 4.03 & 3.84 & 3.87 \\ 
		\hline
		r & 0.46 & 0.43 & 0.42 & 0.40  & 0.38  & 0.37 & 0.36 & 0.35 & 0.34 & 0.33& 0.32 \\ 
		\hline
	\end{tabular}
	\caption{Estimated coefficients by nonlinear regression for fixed $\zeta=1$ and $p=0$}
	\label{tab:RW01}
\end{table}

\begin{table}[h!]
	\centering
	\begin{tabular}{ |c|c|c|c|c|c|c|c|c|c|c|c|} 
		\hline
		$\zeta$ & 1 & 1.1 & 1.2 & 1.3   & 1.4 & 1.5 & 1.6 & 1.7 & 1.8 & 1.9 & 2 \\ 
		\hline
		c & 5.64 & 7.23 & 10.67 & 12.11   & 13.86 & 17.48 & 15.47 & 17.46 & 14.28 & 13.96 & 12.06 \\ 
		\hline
		r & 0.46 & 0.54 & 0.65 & 0.71   & 0.77 & 0.85 & 0.86 & 0.92 & 0.92 & 0.94 & 0.93 \\  
		\hline
	\end{tabular}
	\caption{Estimated coefficients by nonlinear regression for fixed $\lambda=\frac{1}{2}$ and $p=0$}
	\label{tab:RW02}
\end{table}

\begin{figure}[h!]
	\centering
	\subfigure[$\lambda=\frac{1}{2}$]{%
		\label{fig:RW21a}%
		\includegraphics[width=2.8in,height=2.5in]{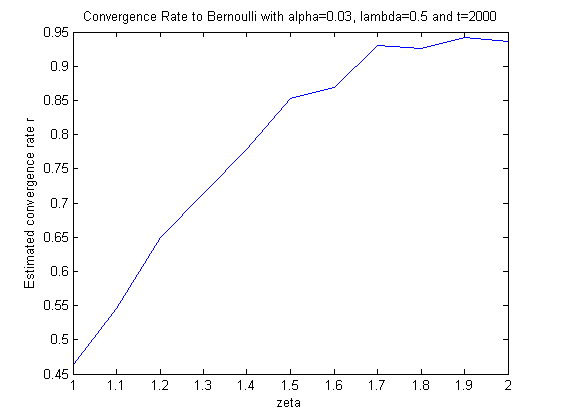}}%
	\hfill
	\subfigure[$\zeta=1$]{%
		\label{fig:RW21b}%
		\includegraphics[width=2.8in,height=2.5in]{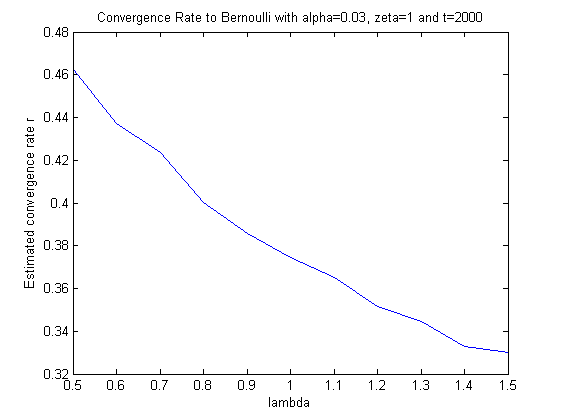}}%	

	\caption{Estimated convergence rates by rational decay model for $p=0$}
	\label{fig:RW21}
\end{figure}

Figures \ref{fig:RW21a} and \ref{fig:RW21b} illustrate the convergent rates for these two situations, i.e. first fixing $\zeta=1$ and varying $\lambda$, and then, varying $\zeta$ and fixing $\lambda=\frac{1}{2}$. Therefore, we observe from nonlinear regression results that when we fix $\zeta=1$, the convergent rate $r$ in function of $\zeta$ is increasing which means that the time-inhomogeneous quantum walk converges faster to Bernoulli distribution while $\zeta$ increases. On the other hand, with fixed $\lambda=\frac{1}{2}$ the convergent rate $r$ in function of $\lambda$ is decreasing which means that it converges slower to Bernoulli distribution if $\lambda$ increases.

Intuitively, if $\lambda,\zeta>0$ the transition matrix
$$
C_n=\begin{bmatrix} 
\sqrt{1-\frac{\lambda}{n^{\zeta}}} & \sqrt{\frac{\lambda}{n^{\zeta}}} \\
\sqrt{\frac{\lambda}{n^{\zeta}}} & -\sqrt{1-\frac{\lambda}{n^{\zeta}}} 
\end{bmatrix}\rightarrow 
\begin{bmatrix} 
1 & 0 \\
0 & 1
\end{bmatrix},
$$
and the rate the matrix converges depends on the term $\frac{\lambda}{n^\zeta}$ in the matrix. Indeed, it's clear that for large $\lambda$'s and  small $\zeta$'s the matrix converges slower than small $\lambda$'s and large $\zeta$'s which accords to our simulation results.

\section{Comparison with decoherent quantum walk when $p=1$ and $\zeta\geq1$}\label{sec:quantumBernoulli}
After we study numerically the convergent rates of the quantum analogues of the classical distributions in the previous section, it is also interesting to compare the obtained results with the convergent rates of the decoherent walks with $p=1$, their classical analogues. In order to do this comparison, we first fit nonlinear regression model to our model with $p=1$ and $\zeta>1$.

\subsection{Convergent rates of the decoherent walk when $\zeta>1$ to Bernoulli }
Note that for $p=1$, the densities converge to Bernoulli distribution with scaling exponent $\gamma=1$, which says that

$$\hat{p}(x,t)\equiv p_d(tx,t),\ x\in \frac{\mathbb{Z}}{t},$$
converges to Bernoulli distribution (See \cite{EnglanderClassic}). Therefore, in this section, we study statistically the convergent rate of them. 

Let us consider $\alpha_t$ as we defined in Section \ref{QuantumToBernoulli}, we fit two nonlinear regression: exponential decay model $ce^{-rt}$ and rational decay model $ct^{-r}$ in Matlab with the same initial value formula we considered,

$$[c_0,r_0]=[1,\frac{\ln(\alpha_{2000})}{2000}],[1,\frac{\ln(\alpha_{2000})}{\ln(2000)}],$$
for exponential and rational models respectively.

\begin{figure}[h!]
	\centering
	\subfigure[$\zeta=\frac{5}{4}$]{%
		\label{fig:RW22a}%
		\includegraphics[width=2.8in,height=2.3in]{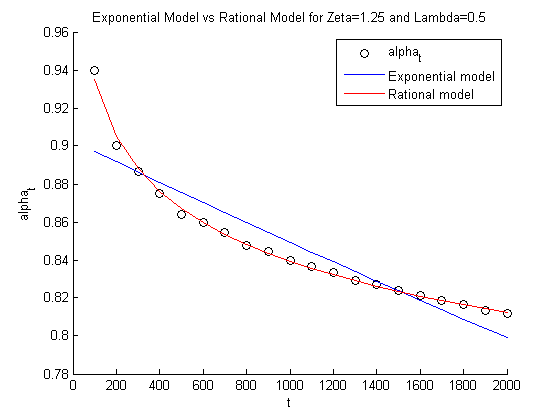}}%
	\hfill
	\subfigure[$\zeta=\frac{9}{5}$]{%
		\label{fig:RW22b}%
		\includegraphics[width=2.8in,height=2.3in]{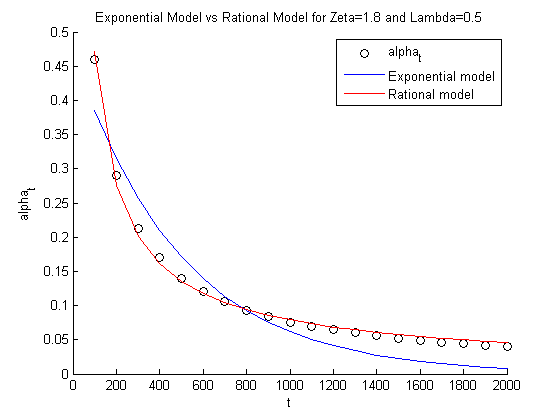}}%	

	\caption{Non linear regression model comparison for $\lambda=\frac{1}{2}$ and $p=1$}
	\label{fig:RW22}
\end{figure}

As results, rational decay model has better R-squared estimate and root mean squared error again. For instance, 

with $\zeta=\frac{5}{4}$ and $\lambda=\frac{1}{2}$, 
the R-squared estimate of the rational decay model is $0.997$, while the R-squared estimate is $0.85$ for the exponential decay model. The root mean error: $0.00197$ for the rational decay model and $0.0131$ for the exponential decay model.

Moreover, Figure \ref{fig:RW22} show both fitted model and $\alpha_t$ and we observe that the rational decay model fits better . It is also clear that the functions graphed by rational model using estimated coefficients by nonlinear regression behave more similarly than the functions graphed by exponential models.

\begin{table}[h!]
	\centering
	\begin{tabular}{ |c|c|c|c|c|c|c|c|c|c|c|c| } 
		\hline
		$\lambda$ & 0.5 & 0.6 & 0.7 & 0.8  & 0.9 & 1 & 1.1 & 1.2 & 1.3 & 1.4 & 1.5 \\ 
		\hline
		c & 3.31 & 2.91 & 2.54 & 2.31  & 2.14  & 1.97 & 1.89 & 1.80 & 1.70 & 1.67 & 1.61 \\ 
		\hline
		r & 0.30 & 0.26 & 0.22 & 0.20  & 0.18  & 0.16 & 0.15 & 0.14 & 0.13 & 0.12 & 0.11 \\ 
		\hline
	\end{tabular}
	\caption{Estimated coefficients by nonlinear regression for fixed $\zeta=\frac{3}{2} $ and $p=1$}
	\label{tab:RW03}
\end{table}

\begin{table}[h!]
	\centering
	\begin{tabular}{ |c|c|c|c|c|c|c|c|c|c|c|c|} 
		\hline
		$\zeta$ & 1.25 & 1.35 & 1.45 & 1.55   & 1.65 & 1.75 & 1.85 & 1.95 & 2.05 & 2.15 & 2.25 \\ 
		\hline
		c & 1.15 & 1.46 & 2.39 & 4.82   & 8.92 & 14.35 & 18.27 & 18.21 & 19.74 & 13.96 & 10.19 \\ 
		\hline
		r & 0.05 & 0.10 & 0.22 & 0.39   & 0.56 & 0.71 & 0.82 & 0.89 & 0.96 & 0.94& 0.93 \\  
		\hline
	\end{tabular}
	\caption{Estimated coefficients by nonlinear regression for fixed $\lambda=\frac{1}{2}$ and $p=1$}
	\label{tab:RW04}
\end{table}

In order to compare the convergence rates with the pure quantum case in the previous section, we execute nonlinear regression for the rational model by fixing $\zeta=\frac{3}{2}$ and varying $\lambda$. then by varying $\zeta$ and fixing $\lambda=\frac{1}{2}$. Tables \ref{tab:RW03} and \ref{tab:RW04} present estimated values of the nonlinear regression for the rational model $\alpha_t\sim ct^{-r}$ we considered.

\begin{figure}[h!]
	\centering
	\subfigure[$\lambda=\frac{1}{2}$]{%
		\label{fig:RW24a}%
		\includegraphics[width=2.7in,height=2.1in]{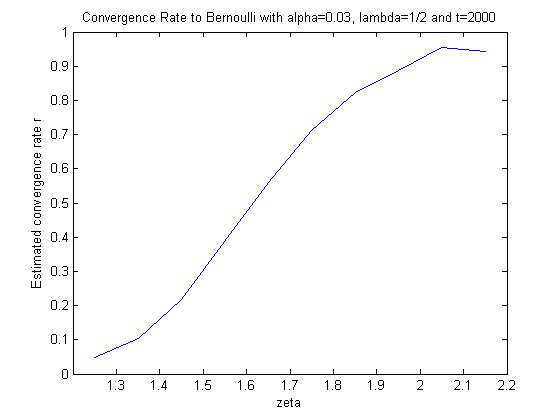}}%
	\hfill
	\subfigure[$\zeta=\frac{3}{2}$]{%
		\label{fig:RW24b}%
		\includegraphics[width=2.7in,height=2.1in]{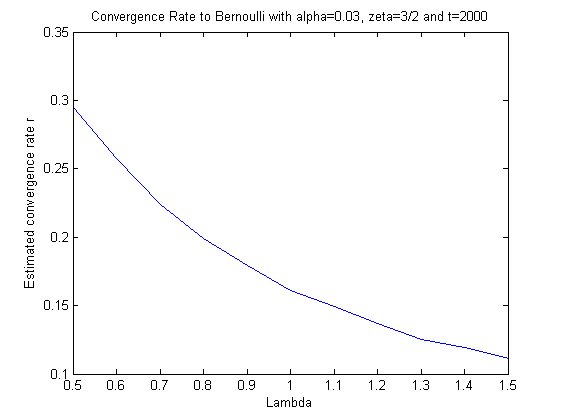}}%
	
	\caption{Estimated convergence rates by rational decay model for $p=1$}
	\label{fig:RW24}
\end{figure}

Figures \ref{fig:RW24a} and \ref{fig:RW24b} illustrate the convergence rates for these two situations, i.e. first fixing $\lambda=\frac{1}{2}$ and varying $\zeta$, and then, varying $\lambda$ and fixing $\zeta=\frac{3}{2}$. Therefore, we observe from nonlinear regression results that when we fix $\zeta=\frac{3}{2}$, the convergence rate $r$ in function of $\zeta$ is decreasing which means that the decoherent quantum walk converges faster to Bernoulli distribution while $\zeta$ increases. On the other hand, with fixed $\lambda=\frac{1}{2}$ the convergence rate $r$ in function of $\lambda$ is increasing which means that it converges slower to Bernoulli distribution if $\lambda$ increases.

The most interesting thing we observe here by comparing the results from the case $p=0$ in previous section is that not only both densities converges to Bernoulli distribution, but also the estimated convergent rates are increasing by fixing $\lambda=\frac{1}{2}$ for both $p=0$ and $p=1$ cases. Moreover, the ranges of estimated convergence rates for $p=0$ and $p=1$ are approximately $[0.45,0.95]$ and $[0.05,0.9]$ respectively for fixed $\lambda=\frac{1}{2}$ and $1<\zeta<2$, which means that the convergence to Bernoulli distribution is faster for the quantum case (see Figures \ref{fig:RW21a} and \ref{fig:RW24a}).

As results, our statistical analysis shows that though our model, not only the quantum analogue of Bernoulli distribution in this case is Bernoulli distributions with the same scaling exponent, but also the convergence speed is even greater that than the decoherent walk when $p=1$.

\subsection{Densities of the decoherent walk when $\zeta=1$}
For this critical case, it is proven that the densities don't converge to Bernoulli distributions. Instead, they converge to symmetric Beta law with parameter $\lambda$, {\bfseries Beta($\lambda$,$\lambda$)}. (see \cite{EnglanderClassic})

\begin{figure}[h!]
	\centering
	\includegraphics[width=1\textwidth]{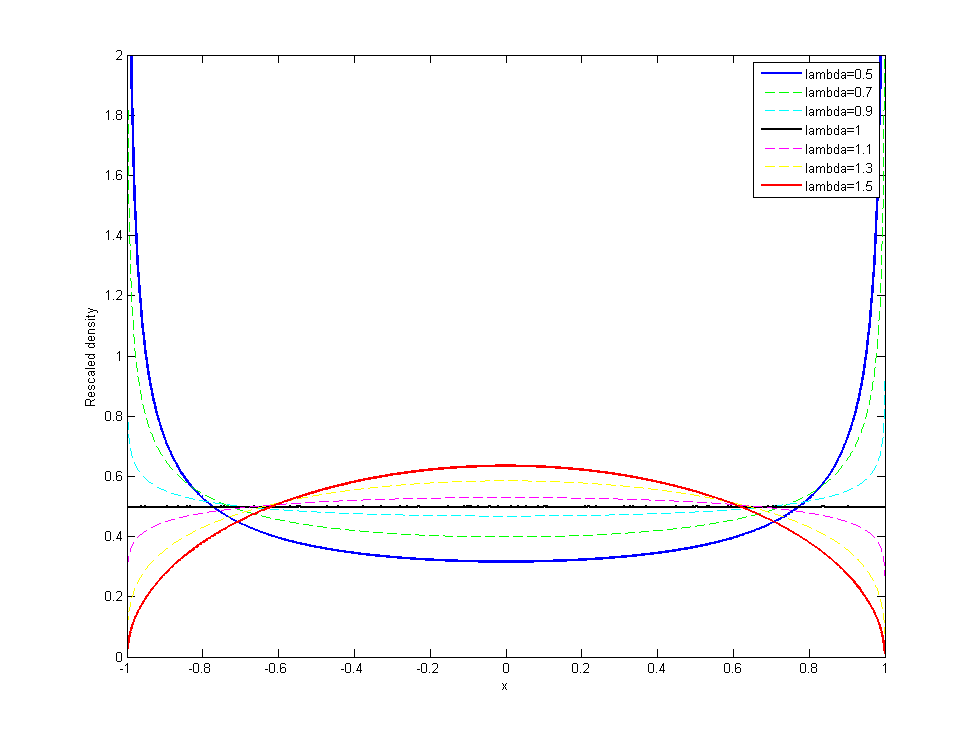}
	\caption{Probability densities with $t=2000$, $\zeta=1$ and $p=1$}
	\label{fig:RW25}
\end{figure}

However, Figure \ref{fig:RW25} illustrates the densities simulated by taking $t=2000$ with different values of $\lambda$'s. We observe that when $\lambda=\frac{1}{2},1,\frac{3}{2}$, the figure shows approximated arcsine, uniform and semicircle densities respectively.

Therefore, this concludes statistically that even though the densities of the decoherent quantum walks for $p=1$ converge to different classical distributions depending the values of $\lambda$, their quantum analogues converge to Bernoulli distributions with different rates of convergence showed in Section \ref{QuantumToBernoulli}. Hence, the quantum analogue of the symmetric Beta distribution in this case is again Bernoulli distribution in the interval $[-1,1]$.

\section{Conclusion}\label{section:conclusion}
In this papaer, we considered the time-inhomogeneous unitary operators, and defined the time-inhomogeneous quantum analogue of the classical random walk with decoherence parameter in $\mathbb{Z}$  and we interpreted the decoherent parameter as the probability to perform a measurement, that means that at each step, we perform a measurement with a certain probability. More specifically, we studied the time-inhomogeneous quantum walk with space-coin decoherence on the set of integers with the time-inhomogeneous unitary operator using the path integral formula and its interpretation as probabilistic geometric measurement time and the time-inhomogeneous Markov chain on the coin space. 

As results, we proved a representation theorem which not only gives us a better probabilistic illustration about the probability at each state, but also gives an approach to approximate the classical distribution through quantum algorithms and a tool to calculate probability densities of quantum walk with decoherence in general.

Additionally, quantized arcsine, uniform, semicircle and Bernoulli distributions were introduced by considering the time-inhomogeneous quantum walk without decoherence on the infinite discrete space. We analyzed their scaling limits and convergence rates statistically using nonlinear regression model, and concluded that not only they all converge to Bernoulli distribution with scaling exponent 1, but also the convergence speeds are higher than their classical analogues.


\begin{thebibliography}{99}
	
\bibitem{AharonovQuanGraph} D. Aharonov and A. Ambainis and J. Kempe and U. Vazirani, Quantum walks on graphs, Proceedings of the 33rd annual ACM symposium on Theory of computing, 50-59, (2001).

\bibitem{BroadbentPerco} S. Broadbent AND J. Hammerslfy, Percolation processes: I and II, Cambridge Philosophical Society, 53, 1957, 629-645, (1957).

\bibitem{BrunDecoh} T. Brun and H. Carteret and A. Ambainis, Quantum random walks with decoherent coins, Phys. Rev., A 67, 032304, (2003).

\bibitem{ChouQMarkov} C. Chou and W. Yang, Time-inhomogendous quantum Markov chains with decoherence on finite state spaces, Arxiv Preprint:2012.05449, (2020).


\bibitem{DietzInhomoMarkov} Z. Dietz and S. Sethuraman, Occupation laws for some time-nonhomogeneous Markov chains. Electron. J. Probab. 12(23), 661–683 (2007)

\bibitem{DobrushinInhomoMarkov} R. Dobrushin: Central limit theorems for non-stationary Markov chains. I., II. Theory Probab. Appl. 1, 65–80 (1956)


\bibitem{Rurrett} R. Durrett, Probability: theory and examples, 5th edition, Cambridge, (2019).

\bibitem{EnglanderClassic} J. Englander AND S. Volkov, Turning a coin over instead of tossing it, Journal of Theoretical Probability, 31, 1097-1118, (2018).

\bibitem{YangConvexNormal} S. Fan and Z. Feng and S. Xiong and W. Yang, Convergence of quantum random walks with decoherence, Phys. Rev., A 84, 042317, (2011).

\bibitem{GantertAnnealing} N. Gantert, Laws of large numbers for the annealing algorithm. Stoch. Process. Appl. 35, 309–313
(1990)


\bibitem{Grover} L. Grover, A fast quantum mechanical algorithm for database search, Proceeding of 28th annual ACM symposium Theory of computing, 212-219, (1996).

\bibitem{Hastings} W. Hastings, Monte Carlo Sampling methods using Markov chains and their applications, Biometrika 57 1, 97-109,(1970).

\bibitem{NonLinearTools} S. Huet and A. Bouvier and M. Poursat and E. Jolivet, Statistical tools for nonlinear regression: a practical guide with S-PLUS and R examples, Springer, (2004).



\bibitem{KirkpatrickAnnealing} S. Kirkpatrick, C. Gelatt, M. Vecchi, Optimization by simulated annealing, Science 220:671–680, (1989).


\bibitem{KonnoHadamardLimit} N. Konno, A new type of limit theorems for the one-dimensional quantum random walk, J. Math. Soc. Japan, 57, 4, 1179-1195, (2005).

\bibitem{LagroQuantumLimit} M. Lagro and W. Yang, A Perron–Frobenius Type of Theorem for Quantum Operations, Journal of Statistical Physics, 169, 38-62, (2017).


\bibitem{Mixing} D. Levin and Y. Peres and E. Wilmer, Markov Chains and Mixing Times, AMS, (2008).


\bibitem{MetroUlam} N. Metropolis and S. Ulam, The Monte Carlo method, Journal of the American statistical association, 44 247, 124-134, (1949).

\bibitem{NonLinearStats} G. Seber and C. Wild, Nonlinear regression, Wiley, (2003).

\bibitem{Shor} P. Shor, Algorithms for quantum computation: Discrete logarithms and factoring, Foundations of Computer Science, 1994 Proceedings 35th Annual Symposium, 335-341, (1994).

\bibitem{SzegedyMarkovC} M. Szegedy, Quantum speed-up of Markov chain based algorithms, Foundations of Computer Science Proceedings 45th Annual IEEE Symposium, 32-41, (2004).


\bibitem{SethuramanInhomoMarkov} S. Sethuraman and S. Varadhan, A martingale proof of Dobrushin’s theorem for non-homogeneous Markov chains. Electron. J. Probab. 10(36), 1221–1235 (2005)

\bibitem{ZehDecoh} H. Zeh, On the Interpretation of Measurement in Quantum Theory, Foundations of Physics, 1, 69-79, (1970).

\bibitem{ZhangQuantumLimit} K. Zhang, Limiting distribution of decoherent quantum random walks, Phys. Rev., A 77, 062302, (2008).



\end{thebibliography}
\end{document}